\documentclass[pdflatex,sn-mathphys-num]{sn-jnl}

\usepackage{graphicx}
\usepackage{xcolor}
\usepackage{bm}
\usepackage{amsmath}
\usepackage{hyperref}
\newcommand{\removeaffilperiod}[1]{}
\usepackage{pdfpages}

\begin{document}

\title{Gate-tunable negative differential resistance in multifunctional
van der Waals heterostructure}

\author*{\fnm{Richa} \sur{Mitra}\textsuperscript{1,\textsection}}
\email{richa@chalmers.se}

\author[2,3]{\fnm{Konstantina} \sur{Iordanidou}}
\author[1]{\fnm{Naveen} \sur{Shetty}}
\author[1]{\fnm{Md Anamul} \sur{Hoque}}
\author[4,5]{\fnm{Anushree} \sur{Datta}}
\author[1]{\fnm{Alexei} \sur{Kalaboukhov}}
\author[2]{\fnm{Julia} \sur{Wiktor}}
\author[1]{\fnm{Sergey} \sur{Kubatkin}}

\author*[1]{\fnm{Saroj Prasad} \sur{Dash}}
\email{saroj.dash@chalmers.se}

\author*[1,6]{\fnm{Samuel} \sur{Lara-Avila}}
\email{samuel.lara@chalmers.se}

\affil[1]{\orgdiv{Department of Microtechnology and Nanoscience}, \orgname{Chalmers University of Technology}, \orgaddress{\postcode{SE-412 96}, \city{Gothenburg}, \country{Sweden}}\removeaffilperiod}

\affil[2]{\orgdiv{Department of Physics}, \orgname{Chalmers University of Technology},
\orgaddress{\postcode{SE-412 96}, \city{Gothenburg}, \country{Sweden}} \removeaffilperiod}

\affil[3]{\orgdiv{Materials Physics-Oslo}, \orgname{SINTEF Industry}, \orgaddress{\postcode{NO-0314}, \city{Oslo}, \country{Norway}} \removeaffilperiod}

\affil[4]{\orgdiv{Laboratoire Matériaux et Phénomènes Quantiques}, \orgname{Université Paris-Cité, CNRS}, \orgaddress{\postcode{75013}, \city{Paris}, \country{France}} \removeaffilperiod}

\affil[5]{\orgdiv{Laboratoire de Physique des Solides}, \orgname{Université Paris-Saclay, CNRS}, \orgaddress{\postcode{91405}, \city{Orsay}, \country{France}} \removeaffilperiod}

\affil[6]{\orgname{National Physical Laboratory}, \orgaddress{\street{Hampton Road}, \city{Teddington}, \postcode{TW11 0LW}, \country{United Kingdom}} \removeaffilperiod}

\abstract{
Two-dimensional (2D) semiconductors have emerged as exciting candidates for the development of low-power and multifunctional computing applications, thanks to their qualities such as layer-dependent band gap tunability, high carrier mobility, and excellent electrostatic control. Here, we explore a pair of 2D semiconductors with nearly broken-gap (Type-III-like) band alignment and demonstrate a highly gate-tunable p-MoTe$_2$/n-SnS$_2$ heterojunction with multifunctional behavior. Employing a dual-gated asymmetric device geometry, we unveil its functionality as both a forward and backward rectifying device. Moreover, we observe a highly gate-tunable negative differential resistance (NDR), with a gate-coupling efficiency of $\eta \simeq 0.5$ and a peak-to-valley ratio of $\sim$ 3 down to 150K. By employing density functional theory, we determine that the observed NDR is dominated by valence band-to-valence band tunneling, while additional interband tunneling contributions arise at higher bias. The combination of tunneling driven transport and gate controllability of NDR opens the pathway for realizing gate-tunable 2D material-based neuromorphic and energy-efficient electronics.
}


\maketitle

\begingroup
\renewcommand{\thefootnote}{\textsection}
\footnotetext{Present address: VTT Technical Research Centre of Finland Ltd.,
Tietotie 3, 02150 Espoo, Finland.}
\endgroup

\section*{Introduction}

The quest for enhanced computational power, energy efficiency, and data processing capabilities continues to fuel the demand for novel materials and computational paradigms \cite{liu2020two, liu2021promises, hoque2024all, kanungo20222d}. Tunnel field effect transistors (TFETs) emerge as exciting candidates to realize the long-standing promise of energy-efficient computation, using band-to-band tunneling (BTBT) \cite{ionescu2011tunnel,boucart2007double,pawlik2012benchmarking, nakamura2020all,kim2020thickness,nourbakhsh2016transport} as the primary carrier injection mechanism, as opposed to conventional thermal injection of field-effect transistors (FETs). This would allow to circumvent inherent fundamental constraints of FETs, because BTBT allows for sharper turn ON/OFF of the transistors and reduce sub-threshold swing (SS = ($\frac{\partial V_G}{\partial I_{SD}}$) below the thermionic limit of 60 mVdec$^{-1}$ at room temperature \cite{
kanungo20222d, ionescu2011tunnel, boucart2007double, pawlik2012benchmarking, nakamura2020all, kim2020thickness, nourbakhsh2016transport} (with I$_\mathrm{SD}$ the source-drain current, and V$_\mathrm{G}$ the gate voltage). Importantly, TFETs can exhibit negative differential resistance (NDR) under specific conditions \cite{shim2016phosphorene,huo2024negative,xiong2020transverse,srivastava2021resonant,yan2015esaki,movva2016room,kim2024room,yan2017tunable,roy2016tunneling,fan2019tunable,xiao2024multifunctional,kinoshita2024negative,britnell2013resonant,fallahazad2015gate,zhang2023toward,seo2022van}, which opens avenues for the implementation of multi-valued logic in emerging applications such as high frequency oscillators \cite{bockle2021gate}, artificial intelligence \cite{han2025progress} and neuromorphic computing \cite{chaves2020bandgap}. 

In recent years, two-dimensional (2D) materials \cite{liu2020two, liu2021promises, hoque2024all, kanungo20222d} are being explored to boost the performance of TFETs and to bypass challenges often encountered when implementing TFETs with bulk semiconductors: the absence of dangling bonds on 2D crystals ensures a pristine, bond-free surface, and allows for the creation of atomically clean and sharp interfaces in 2D van der Waals (vdW) heterostructures. Moreover, 2D material TFETs provide an edge over their three-dimensional counterparts, as reduced screening in two dimensions facilitates better electrostatic control of devices. With the plethora of 2D materials available at our disposal, it is crucial to carefully select combinations of materials that act as source and the channel of the TFET. A common design rule is to choose a narrow bandgap p-type source and a wider bandgap n-type channel. This enhances the density of tunneling carriers, enabling efficient carrier injections from the source to the channel. The larger bandgap of the channel material suppresses off-state leakage, thereby improving the device’s ON/OFF current ratio. Several heterojunctions implemented following these prescriptions have been demonstrated \cite{boucart2007double,shim2016phosphorene,huo2024negative,xiong2020transverse,srivastava2021resonant,yan2015esaki,movva2016room,kim2024room,yan2017tunable,roy2016tunneling,fan2019tunable,xiao2024multifunctional,kinoshita2024negative}. Recent theoretical works have sparked interest into unconventional pairs of 2D semiconductors \cite{chaves2020bandgap,bao2018band,iordanidou2023electric} having broken (Type-III alignment) or nearly broken-gap band alignment, suitable for TFET operation, as it facilitates efficient band-to-band-tunneling, that would result in high on-state current, low subthreshold swing, reduced power consumption and faster logic operations.

Here we explore one such unconventional pair of 2D semiconductors in the light of tunable nearly broken-gap (Type-III-like) band alignment by combining a few layers of p-MoTe$_2$ and n-SnS$_2$, which are predicted to form a naturally nearly broken-gap band alignment \cite{iordanidou2022two,guo2023tunable,liang2025computational}. We demonstrate multifunctional behavior of the junction, showing tunable forward to reverse rectifying behavior with gate voltage and observe gate-tunable NDR \cite{huo2024negative,kim2024room,roy2016tunneling,fan2019tunable,seo2022van,bockle2021gate,han2025progress, kim2020gate,chen2022reversible} at temperatures down to 150K. Our first-principles calculations based on density functional theory (DFT) shed light on our experimental observations and provide insight in device optimization pathways. 

\section*{Results: theoretical prediction}

\begin{figure*}[t]
    \centering
    \includegraphics[width=0.9\textwidth]{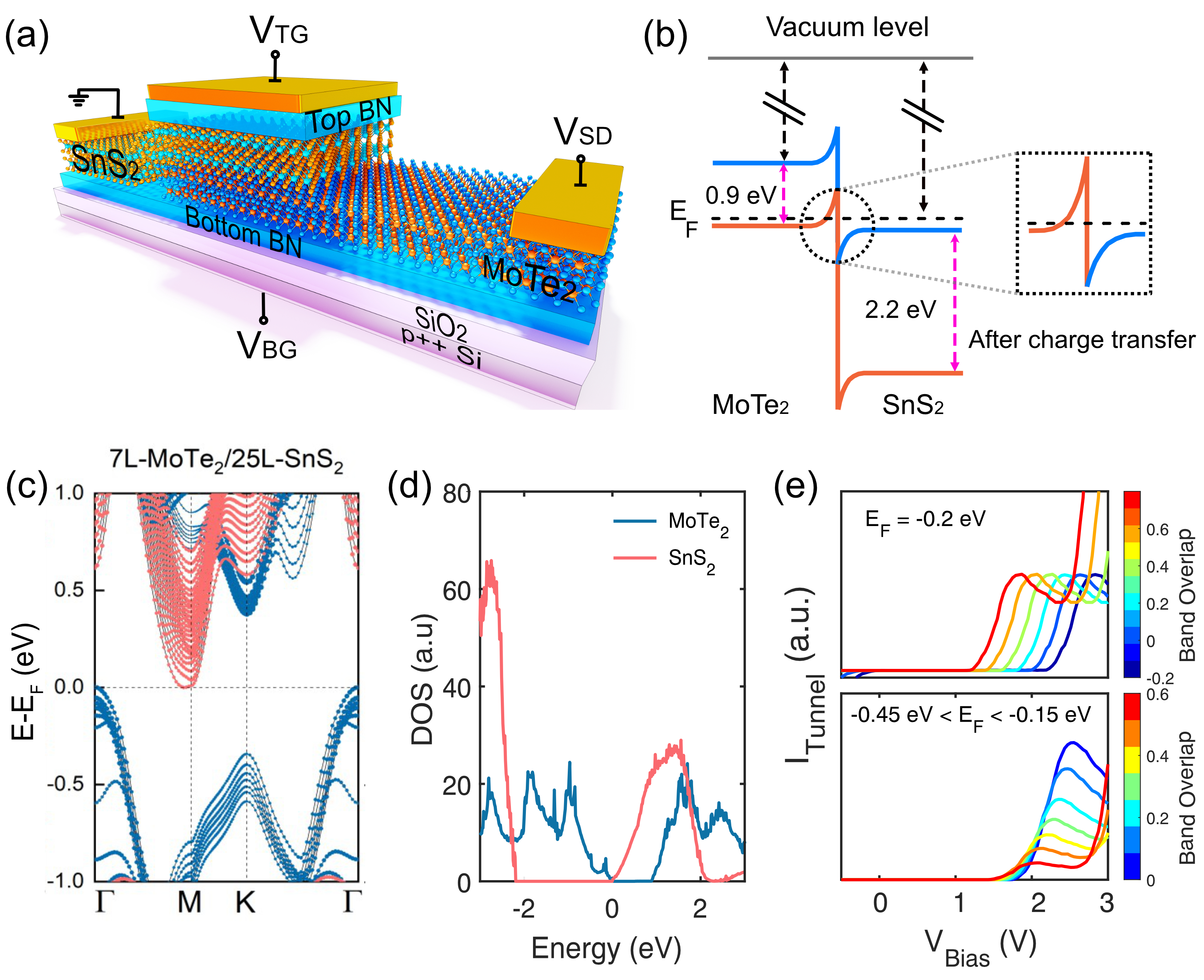}
    \caption{\textbf{Type-III heterojunction with MoTe$_2$/SnS$_2$: (a)} Device schematic of a dual gated vertical heterojunction of MoTe$_2$ and SnS$_2$ on a Si/SiO$_2$ substrate. Backgate V$_\mathrm{BG}$ and topgate voltages V$_\mathrm{TG}$ are applied across the SiO$_2$ substrate and top hBN respectively to change the carrier density. Biasing configuration is also shown, where V$_\mathrm{SD}$ is applied to MoTe$_2$, SnS$_2$ is grounded. \textbf{(b)} Nearly broken-gap (type-III-like) band alignment of MoTe$_2$ and SnS$_2$ after forming the junction. Band bending appears as a result of charge transfer from MoTe$_2$ to SnS$_2$ (KPFM results in Supplementary Info section S4). The enlarged view at the interface illustrates the nearly broken-gap alignment with a small overlap between the MoTe$_2$ valence band and SnS$_2$ conduction band. \textbf{(c)} Band structure from DFT calculations for 7-layer MoTe$_2$/25-layer SnS$_2$ heterojunction showing MoTe$_2$ VB maxima aligning with SnS$_2$ CB minima. \textbf{(d)} Partial density of states of MoTe$_2$ and SnS$_2$ from DFT. \textbf{(e)} Tunnel current (I$_\mathrm{Tunnel}$) calculated from equation \ref{eq1} using DFT DOS, plotted w.r.t theoretically applied bias voltage (V$_\mathrm{Bias}$). I$_\mathrm{Tunnel}$ are estimated using DFT DOS for different band alignment configurations: (Upper panel) The extent of band overlaps i.e., the broken-gap alignment is varied (as indicated by the color bar, in units of eV), while the Fermi energies are held constant across the junction (E$_\mathrm{Fm}$ = E$_\mathrm{Fs}$ = -2 eV). In this configuration, I$_\mathrm{Tunnel}$ remains unchanged while V$_\mathrm{peak}$ shifts systematically. (Bottom panel) Simultaneously varying the Fermi energies (e.g. -0.45 eV $<$  E$_\mathrm{F} < $ -0.15 eV) and the band overlap results in systematic changes in both the magnitude of I$_\mathrm{Tunnel}$ and the position of V$_\mathrm{peak}$ (SI section S2).}
    \label{fig:fig1}
\end{figure*}

A schematic of our proposed device, a MoTe$_2$/SnS$_2$ heterojunction, is shown in Fig. \ref{fig:fig1}(a). In this device architecture, MoTe$_2$ has bandgap around $\approx$ 0.9 eV and acts like a source material, whereas SnS$_2$ having a larger bandgap ($\approx$ 2.2 eV) and higher electron affinity serves as a channel material.  Figure \ref{fig:fig1}(b) illustrates the schematic of a nearly broken-gap band alignment in vacuum. A qualitative schematic illustrating charge-transfer-induced band bending at the interface has been shown in the Supplementary Information (Fig. S4).  To support the band alignment, we performed first-principles DFT calculations with van der Waals corrections, which reveal a nearly broken-gap configuration at the MoTe$_2$/SnS$_2$ interface. Figure \ref{fig:fig1}(c) shows the corresponding DFT-computed band structure, highlighting the overlap between the MoTe$_2$ valence band (VB) and the SnS$_2$ conduction band (CB) density of states. While Fig. \ref{fig:fig1}(c) presents the results for a specific layer thickness combination, our calculations \cite{iordanidou2023electric,iordanidou2022two,liang2025computational} show that the nearly broken-gap alignment is a robust feature across various MoTe$_2$/SnS$_2$ heterostructure thicknesses (see Supplementary info SI Section S1). The first-principles calculations also allow us to anticipate the device performance of the MoTe$_2$/SnS$_2$ heterojunction. Specifically, we use the partial density of states (DOS) of MoTe$_2$ and SnS$_2$ to compute the tunneling current using the following expression \cite{shim2016phosphorene}:

\begin{equation}
I_{\mathrm{Tunnel}}
=
\int_{E_{\mathrm{Fm}}}^{E_{\mathrm{Fs}}}
\mathrm{DOS}_{\mathrm{MoTe_2}}(E)
\mathrm{DOS}_{\mathrm{SnS_2}}(E-eV_{\mathrm{Bias}})
\left[
f_{\mathrm{MoTe_2}}(E)
-
f_{\mathrm{SnS_2}}(E-eV_{\mathrm{Bias}})
\right]
\,\mathrm{d}E
\label{eq:tunnel}
\end{equation}

Here $\mathrm{DOS}_{\mathrm{MoTe_2}}$ ($\mathrm{DOS}_{\mathrm{SnS_2}}$) and $f_{\mathrm{MoTe_2}}$ ($f_{\mathrm{SnS_2}}$) represent the DOS and the Fermi distribution function of MoTe$_2$ (SnS$_2$) respectively. In Fig. \ref{fig:fig1}(d) we plot the partial DOS of the two materials with zero ‘band overlap’ which indicates that the VB maxima of MoTe$_2$ aligns with CB minima of SnS$_2$ at Energy = 0. We define negative (positive) ‘band overlap’ to the scenario when SnS$_2$ CB minima lie lower (higher) in energy than VB maxima of MoTe$_2$. We also retain the usual convention of biasing where the positive (negative) V$_{Bias}$ shifts the SnS$_2$ bands towards higher (lower) energy with respect to the MoTe$_2$ bands (detail calculation shown in SI Section S2). 

Our calculations of the theoretical tunneling current ($\mathrm{I}_\mathrm{Tunnel}$) through the MoTe$_2$/SnS$_2$ heterojunction as a function of the bias voltage ($\mathrm{V}_\mathrm{Bias}$) indicate that such device will display tunable NDR features. Figure. \ref{fig:fig1}(e) shows $\mathrm{I}_\mathrm{Tunnel}$ for two different band alignment configurations. The upper panel refers to a scenario where the Fermi levels remain constant across the heterojunction while the extent of the band overlap is varied, transitioning from a partially staggered (Type-II) to a nearly broken-gap configuration (Type-III-like). In this case, the voltage corresponding to the NDR peak ($\mathrm{V}_\mathrm{peak}$) shifts to lower bias values as the band overlap increases. In contrast, under the second configuration where both the Fermi levels and band overlap vary systematically across the junction, both the peak current magnitude and the peak voltage increase. These results indicate that the NDR response in MoTe$_{2}$/SnS$_{2}$ heterojunctions is robust and can be modulated through electrostatic gating. While the nearly broken-gap alignment permits interband VB-CB tunneling in principle, our calculations with overlapping DOS under applied bias indicate that the dominant contribution to the observed NDR originates from tunneling between valence band states of MoTe$_{2}$ and SnS$_{2}$, as discussed further in Supplementary Section S2. We note that the calculations shown in Fig. \ref{fig:fig1}(c) were performed for a specific combination of layer thicknesses. Although thickness dependent variations in the band gap, screening, and work function can quantitatively modify the band overlap, the observed gate-tunable NDR in our MoTe$_{2}$/SnS$_{2}$ devices is consistent with a nearly broken-gap/type-III-like alignment. Direct thickness-resolved measurements of the band offsets would require additional techniques such as KPFM or optical spectroscopy and are beyond the scope of the present work.

\section*{Results: device fabrication}

To fabricate devices, we stack 2D material flakes into heterostructures using a polymer-based hot pickup technique. After exfoliating the flakes onto Si/SiO$_2$ and identifying suitable ones, the polymer-based pick-up at elevated temperatures ensures residue-free interfaces, which is critical for device fabrication. To enhance the interface quality, we perform vacuum annealing followed by an atomic force microscopy (AFM) scan to ensure surface cleanliness. We use Ti (5 nm)/Au (70 nm) and Pd (15 nm)/Au (60 nm) for making ohmic contacts to SnS$_2$ and MoTe$_2$, respectively. This contact engineering approach results in high p-type doping in few-layer MoTe$_2$ and n-type doping in SnS$_2$. For the dual-gated geometry, hBN/MoTe$_2$/SnS$_2$/hBN stacks are fabricated by adding a hBN layer ($\approx$ 20 nm) over the junction. We have fabricated 5 devices according to the above-mentioned process (all fabrication and characterization details in SI section S3, S4). However, we present extensive data for one backgated device (D1) and two dual-gated devices (D2, D3) (shown in Fig. \ref{fig:fig2}(a)). During device fabrication and before electrical measurements we characterize the thickness/morphology and chemical composition of individual flakes using atomic force microscopy (AFM) and Raman spectroscopy, respectively. The AFM topography reveals that MoTe$_2$ and SnS$_2$ flakes are multilayer (in the range of 4-9 layers for the former one and 12-25 layers for the later, respectively as shown in the inset of Fig. \ref{fig:fig2}(a)). Raman analysis in Fig. \ref{fig:fig2}(b) shows typical modes for each layer present in the junction, with pronounced modes at 233 cm$^{-1}$ for MoTe$_2$ and at 315 cm$^{-1}$ for SnS$_2$. 


\subsection*{Backgated device}

\begin{figure*}[t]
    \centering
    \includegraphics[width=1.0\textwidth]{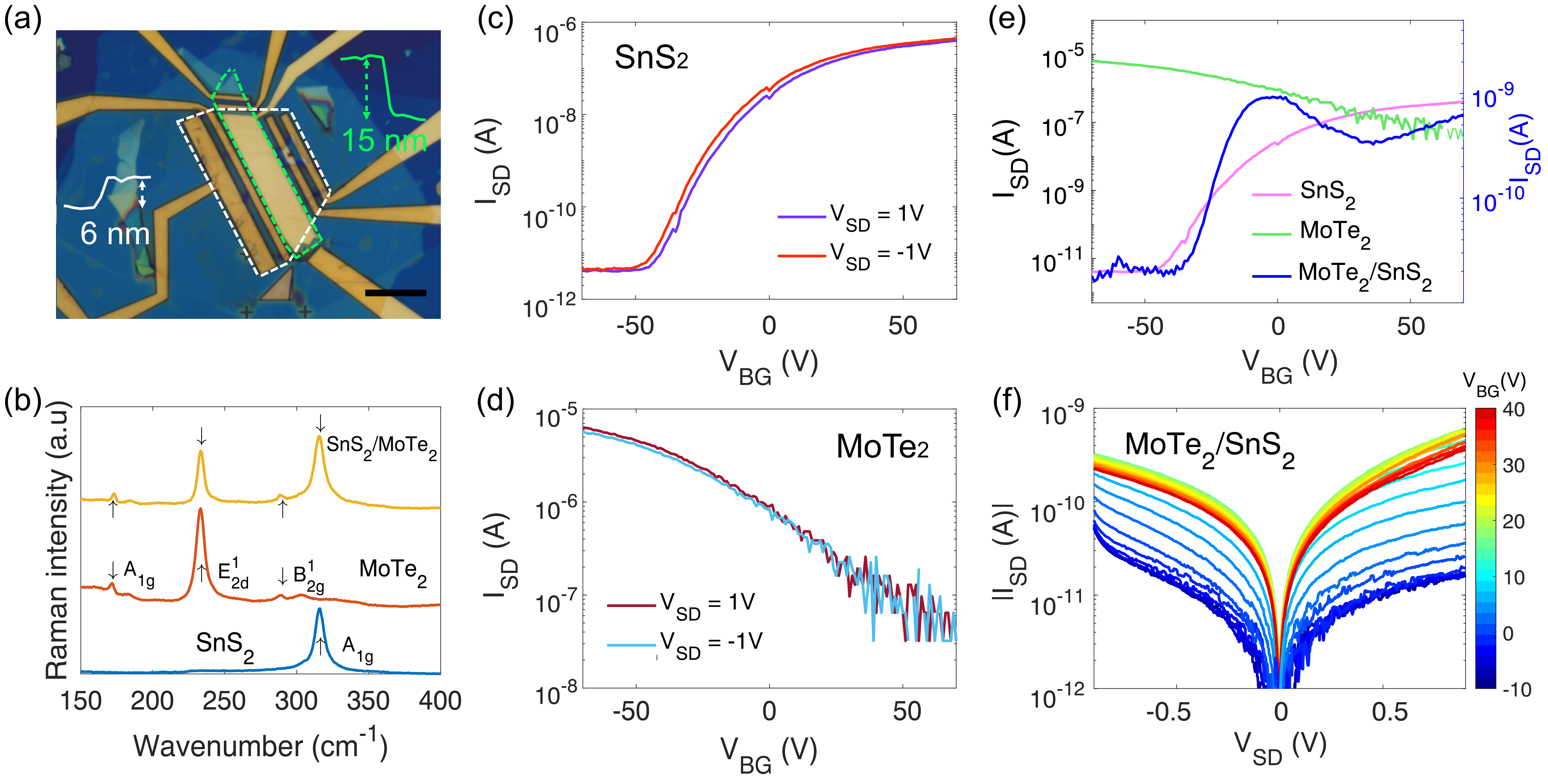}
    \caption{\textbf{Device characterisation from backgated device D1: (a)} Optical micrograph of the fabricated dual-gated MoTe$_2$/SnS$_2$ heterojunction device. The scale bar represents 10 $\mu \mathrm{m}$. The inset shows the AFM height profiles of the MoTe$_2$ and SnS$_2$ flakes. \textbf{(b)} Material characterization of the individual layers using Raman spectroscopy. We used a 532 nm laser source for acquiring the Raman signal of the individual flakes as well as the junction of the device. The Raman spectra show the characteristic A$_{1g}$ phonon mode of SnS$_2$ \cite{sriv2018low} at 315 cm$^{-1}$. For MoTe$_2$ \cite{fraser2020selective}, Raman peaks are observed at wave numbers 233 cm$^{-1}$, 173 cm$^{-1}$, and 289 cm$^{-1}$, which correspond to the in-plane E$^1_{2g}$, out-of-plane A$_{1g}$, and B$^1_{2g}$ modes, respectively. The MoTe$_2$/SnS$_2$ junction carries Raman signals from both materials (indicated by arrows), as shown in the yellow curves. Transfer characteristic curves of SnS$_2$ \textbf{(c)} and MoTe$_2$ \textbf{(d)} measured at room temperature for V$_\mathrm{SD} = \pm$ 1V and plotted in logarithmic scale for backgate voltages from -70V to 70V. \textbf{(e)} Transfer curves of the MoTe$_2$/SnS$_2$ junction (blue, right axis) plotted together with MoTe$_2$ (green, left axis), SnS$_2$ (pink, left axis) for V$_\mathrm{SD}$ = 1V. \textbf{(f)} Output characteristic i.e. I$_\mathrm{SD}$ vs. V$_\mathrm{SD}$ curves for the MoTe$_2$/SnS$_2$ junction for a range of backgate voltages (shown in colorbar) showing modulation of both forward and reverse bias current with V$_\mathrm{BG}$.}
    \label{fig:fig2}
\end{figure*}

To understand the performance of the MoTe$_2$/SnS$_2$ junction, we first present the response of individual flakes for a backgated device D1 at room temperature. Fig. \ref{fig:fig2}(c) and (d) show the transfer characteristics (source-drain current I$_\mathrm{SD}$ vs. backgate voltage V$_\mathrm{BG}$) of SnS$_2$ and MoTe$_2$ flakes respectively at a fixed source-drain voltage, V$_\mathrm{SD} = \pm$ 1V. SnS$_2$ alone exhibits n-type FET behavior with $\sim$ 5 orders magnitude increase in current from OFF to ON state. In comparison, MoTe$_2$ exhibits p-type conduction with high current density; its lower ON/OFF ratio can be attributed to its higher intrinsic doping and to its small band gap, which allows simultaneous transport of both holes and electrons. The nearly identical responses for both bias polarities indicate ohmic contact formation, as further confirmed by the linear I–V characteristics (see SI Section S5A). We further estimate the effective Schottky barrier at the Pd/Au-MoTe$_2$ and Ti/Au-SnS$_2$ contacts from the temperature dependent transport measurements of a representative backgated device to examine the carrier injection at the metal-semiconductor interfaces (see SI Section S5B and Fig. S5B). The extracted barriers provide insight into the contact properties of the individual flakes. Next, Fig. \ref{fig:fig2}(e) shows the transfer characteristics of the individual flakes SnS$_2$ and MoTe$_2$ plotted on the left axis, along with the corresponding response of the MoTe$_2$/SnS$_2$ junction (plotted on the right axis). We find that the current across the junction shows a non-monotonic behavior with respect to the applied backgate voltage, which we attribute to the creation of a p/n density profile along the junction as a result of screening of the applied electric fields. Figure \ref{fig:fig2}(f) completes the device characterization, depicting the I-V characteristic curves of the MoTe$_2$/SnS$_2$ junction in logarithmic scale for a range of backgate voltages. We observe that the source-drain current systematically increases with the applied backgate voltage, indicating a reduction in junction resistance. This increase is evident for both forward and reverse bias, showing no significant rectification behavior. Additionally, the absence of NDR suggests that the junction remains in a regime of limited band overlap and strong electrostatic screening, preventing the formation of a broken-gap alignment necessary for NDR onset \cite{huo2024negative,kim2024room,fan2019tunable,yan2023mote2,kim2020gate}. This highlights the importance of employing dual-gated geometries to enable more effective control over band alignment and carrier distribution across the junction. 

\subsection*{Dual-gated device}

\begin{figure*}[t]
    \centering
    \includegraphics[width=1.0\textwidth]{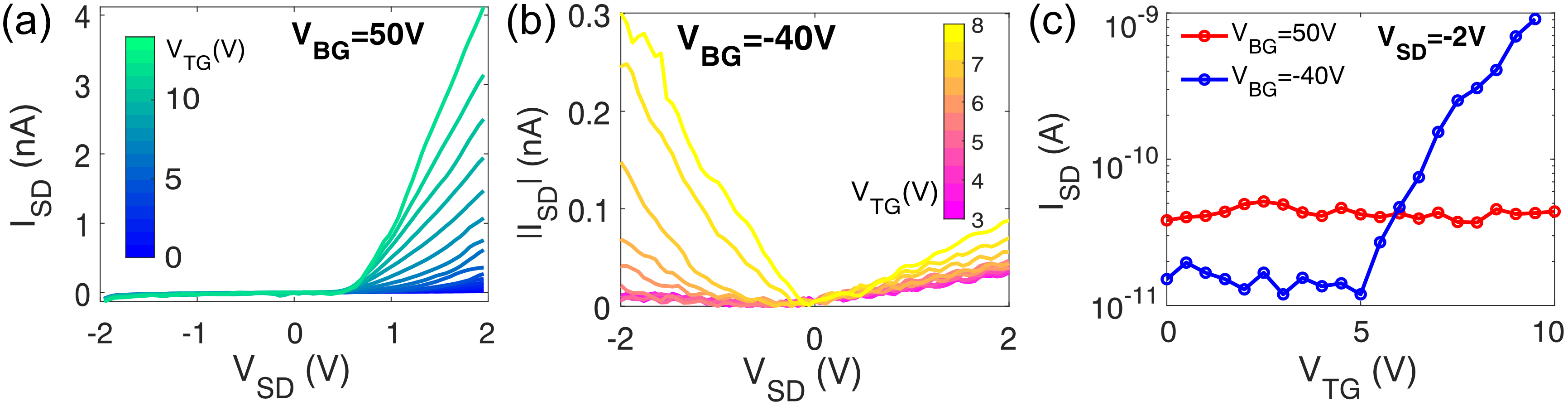}
    \caption{\textbf{Gate-dependent current rectification:} Output curves for opposite polarities of the backgate voltages of \textbf{(a)} 50 V and \textbf{(b)} -40 V showing rectification at forward and reverse bias respectively. \textbf{(c)} Topgate response of the junction current at reverse bias (V$_\mathrm{SD}$ = -2V) comparing rectification at -40V to 50V backgate voltage (Data for V$_\mathrm{SD}$ = 2V is shown in SI section S6).}
    \label{fig:fig3}
\end{figure*}

To enhance the junction tunability and explore its performance as TFET device, we implement a topgate electrode using hBN as the dielectric in addition to the global bottom gate. With the topgate primarily modulating the carrier concentration of the SnS$_2$ flake, our measurements reveal multifunctional transport behavior of the MoTe$_2$/SnS$_2$ heterojunction under opposite backgate polarities \cite{yan2023mote2,kim2020gate}. Figures \ref{fig:fig3}(a) and (b) demonstrate that, at certain combinations of bottom and topgate voltages (V$_\mathrm{TG}$) our device acts as a rectifier in both forward bias (FB) and reverse bias (RB) conditions. In Figs. \ref{fig:fig3}(a) and (b), the I-V curves are plotted on a linear scale for a range of V$_\mathrm{TG}$, as indicated in the colorbars. Figure \ref{fig:fig3}(a) shows at positive backgate voltage (50V) the FB drain current increases significantly with rising V$_\mathrm{TG}$, while RB current remains low and unaffected, because increasing the topgate voltage raises the SnS$_2$ Fermi energy creating more charge carriers for conduction in FB, while the RB current is limited by interband tunneling, which remains suppressed due to insufficient overlap between the DOS. In contrast, Fig. \ref{fig:fig3}(b) shows that for a negative polarity of backgate voltage (-40V), pronounced current rectification is observed for negative source drain bias, arising from charge carriers tunneling from the filled VB of MoTe$_2$ to empty CB band of SnS$_2$ \cite{kim2020gate,chen2022reversible}. We note that the rectification mechanism described in this section and the NDR discussed later corresponds to different backgate regimes and should not be attributed to the same tunneling process. At V$_\mathrm{BG}$ = -40 V, the strong p-doping of MoTe$_2$ favors VB-CB interband tunneling responsible for the enhanced reverse-bias current. In contrast, the NDR observed at V$_\mathrm{BG}$ = -10 V occurs under a different electrostatic band alignment and is predominantly governed by VB-VB tunneling, as supported by our DFT calculations and DOS-overlap analysis, with an additional VB-CB contribution possible at higher bias. Figure \ref{fig:fig3}(c) summarizes the rectification behavior by showing the change of source-drain current with topgate voltage, at different polarities of the backgate voltage for a fixed negative bias (-2V). The observed rectification can be qualitatively understood by examining the transport characteristics of the individual constituent flakes (Figs. \ref{fig:fig2}c and \ref{fig:fig2}d) \cite{ionescu2011tunnel,boucart2007double,seo2022van}. The gate-dependent rectification behavior reflects changes in the relative carrier availability in MoTe$_2$ and SnS$_2$ under different backgate polarities, resulting in suppressed reverse-bias current when MoTe$_2$ is depleted and enhanced tunneling when it is p-doped. A qualitative band-alignment schematic, provided in Supplementary Section S6 (Fig. S6), is included only as a guiding illustration.

\section*{Negative differential resistance}

\begin{figure*}[t]
    \centering
    \includegraphics[width=0.95\textwidth]{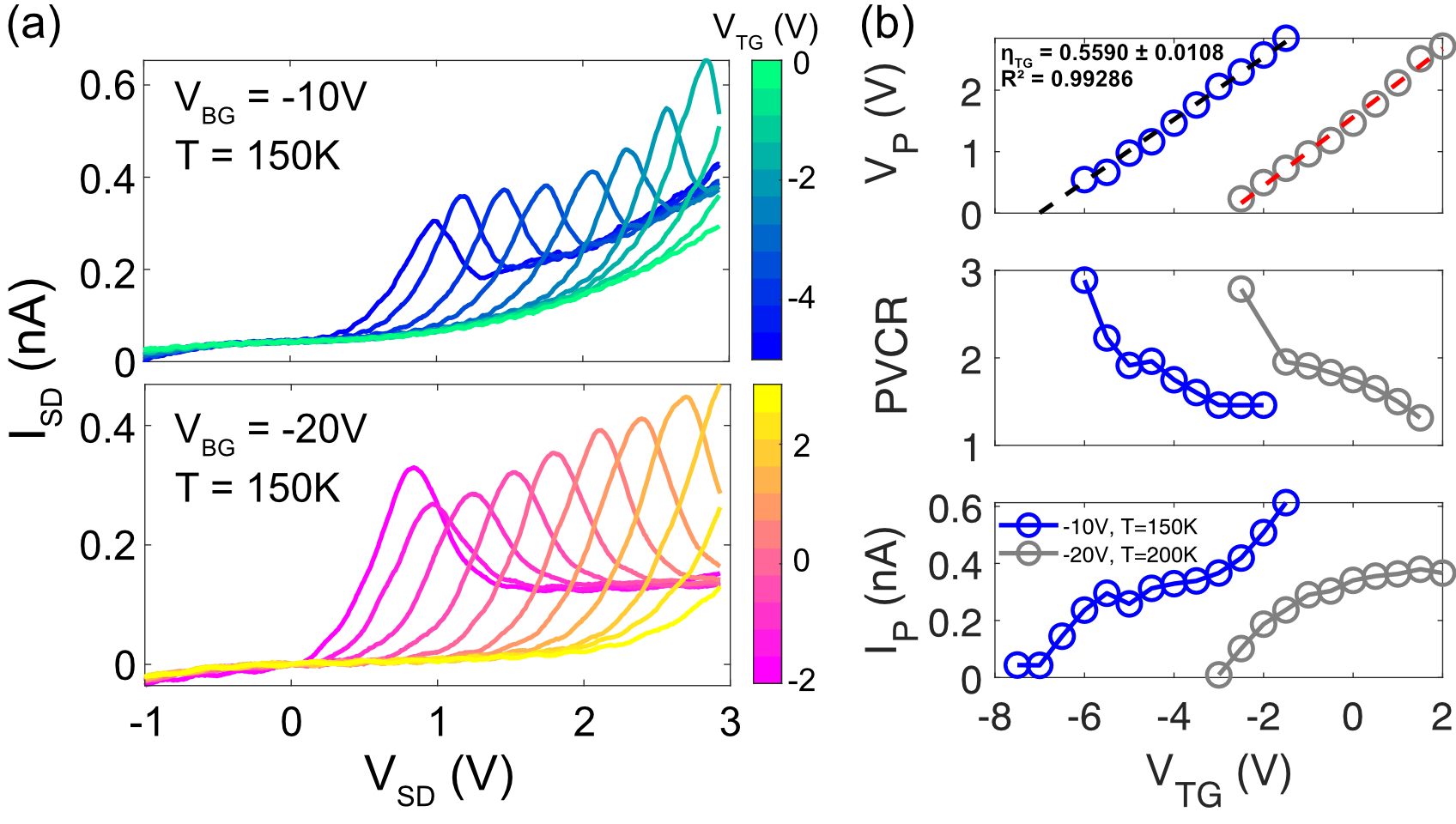}
    \caption{\textbf{Characterization of gate-tunable negative differential resistance in dual gated devices: (a)} Output curves of the dual-gated device D2 showing NDR response at V$_\mathrm{BG}$ = -10V (top panel), at V$_\mathrm{BG}$ = -20V (bottom panel) measured at T=150K. Colorbars represent topgate voltages. \textbf{(b)} (top panel) V$_\mathrm{P}$ (bias voltage at which the NDR peak appears) vs. V$_\mathrm{TG}$ plot. Linear fits yield top-gate coupling efficiencies of $\eta_{\, \mathrm{TG}} = 0.5291 \pm 0.0114$ ($\mathrm{R}^2$ = 0.99633) over V$_\mathrm{TG}$ = -6 to -1.5 V (black dashed line) and $\eta_{\, \mathrm{TG}} =  0.5590 \pm 0.0108$ ($\mathrm{R}^2$ = 0.99286) over V$_\mathrm{TG}$ = -2.5 to 2 V (red dashed line) for the blue and gray datasets, respectively. (Middle panel) Peak-to-valley current ratio (PVCR = I$_\mathrm{Peak}$/I$_\mathrm{Valley}$) vs. V$_\mathrm{TG}$ plot. (Bottom panel) Peak current I$_\mathrm{P}$ vs. V$_\mathrm{TG}$. The blue and the gray circles (curves) represent data from device D2 measured at V$_\mathrm{BG}$ = -10V at 150K and V$_\mathrm{BG}$ = -20V at 200K, respectively. }
    \label{fig:fig4}
\end{figure*}

Having described the gate-tunable rectification in MoTe$_2$/SnS$_2$, we now turn to the observation of NDR in dual-gated devices. Our theoretical model indicates that NDR in the MoTe$_2$/SnS$_2$ heterostructure emerges when the band alignment approaches a nearly broken-gap configuration, where enhanced band overlap leads to efficient valence band-to-valence band tunneling, while additional interband tunneling contributions arise at higher bias. Such a regime is realized in our heterostructure for specific combinations of top- and backgate voltages.

In Fig. \ref{fig:fig4}(a), upper panel, we present I-V curves at a fixed backgate voltage V$_\mathrm{BG}$ = -10V, using the topgate voltage V$_\mathrm{TG}$ as parameter (indicated in the colorbar). We observe that, for each I-V curve the current changes non-linearly with the bias voltage, increasing from negative voltages until it reaches a local maximum at certain bias (peak voltage, V$_\mathrm{P}$).  The effect of the topgate is twofold. First, the topgate allows to tune the position of the NDR peak voltage V$_\mathrm{P}$: as V$_\mathrm{TG}$ reduces from 0V, V$_\mathrm{P}$ systematically shifts to a lower bias voltage. Second, we also observe orderly change in the peak current (= I$_\mathrm{P}$) with the application of topgate voltage. For comparison, the bottom panel of Fig. \ref{fig:fig4}(a) shows similar gate-tunable NDR features are observed for a different backgate voltage, V$_\mathrm{BG}$ = -20V (See SI section S7 for NDR in other experimental conditions) \cite{huo2024negative,kim2024room,roy2016tunneling,fan2019tunable,yan2023mote2,kim2020gate,chen2022reversible}.

To gain more insight into the observed NDR feature, we study the topgate dependence of the transport characteristics of our dual-gated devices. The top panel of Fig. \ref{fig:fig4}(b) shows the NDR peak voltage (V$_\mathrm{P}$) as a function of V$_\mathrm{TG}$ for two representative datasets. In both cases, V$_\mathrm{P}$ varies linearly with V$_\mathrm{TG}$, demonstrating strong top-gate control of the NDR peak position. We define the top-gate coupling efficiency as $\eta_{\, \mathrm{TG}} \, = \, \left| \frac{\Delta V_{P}}{\Delta V_{TG}} \right|$.  For V$_\mathrm{BG}$ = -10 V at T = 150 K, a linear fit over V$_\mathrm{TG}$ = -6 to -1.5 V yields $\eta_{\, \mathrm{TG}} = 0.5291 \pm 0.0114$ with $\mathrm{R}^2$ = 0.99633. For V$_\mathrm{BG}$ = -20 V at T = 200 K, fitting over V$_\mathrm{TG}$ = -2.5 to 2 V gives $\eta_{\, \mathrm{TG}} = 0.5590 \pm 0.0108$ with $\mathrm{R}^2$ = 0.99286. Thus, both datasets consistently yield $\eta_{\, \mathrm{TG}} \approx$ 0.5. This corresponds to an approximately 0.5V shift in the NDR peak for every 1V change in topgate voltage. The high $\eta_{\, \mathrm{TG}}$ indicates strong capacitive coupling and efficient top-gate control of the nearly broken-gap band alignment \cite{kim2020gate,chen2022reversible}. The corresponding back-gate coupling efficiency, $\eta_{\, \mathrm{BG}}$, extracted for device D3, is discussed separately in Section S8 of the Supplementary Information. The considerably lower $\eta_{\, \mathrm{BG}} \approx$ 0.1 compared with $\eta_{\, \mathrm{TG}} \approx$ 0.55 indicates that the top gate provides substantially stronger electrostatic control over the NDR peak position and the corresponding band alignment. Similar gate-induced modulation of the tunneling barrier has been reported in other 2D vdW devices, highlighting the importance of electrostatic control in determining charge injection and transport across such junctions \cite{huo2025bias}.

Figure \ref{fig:fig4}(b) (middle panel) shows the V$_\mathrm{TG}$ dependence of the Peak-to-Valley Current Ratio, PVCR = $\frac{\mathrm{I}_\mathrm{Peak}}{\mathrm{I}_\mathrm{Valley}}$, a key performance indicator of NDR devices which are comparable to or better than previously reported values in other 2D heterostructures \cite{huo2024negative,kim2024room,fan2019tunable,zhang2023toward}. In our device, we find the maximum PVCR to be $\sim$ 3 - 4.  Figure \ref{fig:fig4}(b) (bottom panel) shows the peak current as a function of the topgate voltage plotted for the similar V$_\mathrm{BG}$ and temperature combinations. The observed increase in peak current with topgate voltage indicates enhanced tunneling probability, which likely arises from reduced band misalignment and more efficient carrier injection across the junction. 

\section*{Discussion}

\begin{figure*}[t]
    \centering
    \includegraphics[width=1.0\textwidth]{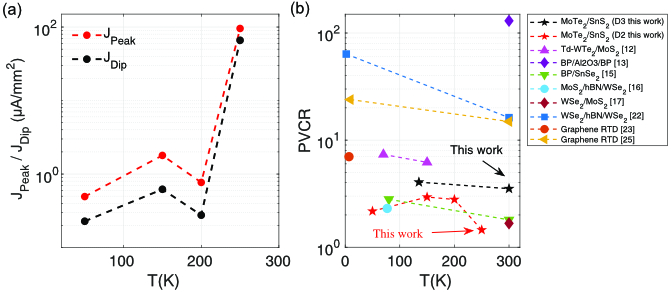}
    \caption{\textbf{Temperature dependence: (a)} Temperature dependence of the peak and dip current densities in our MoTe$_2$/SnS$_2$ devices (normalized by device area 132 $\mu \mathrm{m}^2$ and plotted in units of $\mu$A/mm$^2$ for device D2). \textbf{(b)} Plot of peak-to-valley-ratio (PVCR) vs temperature for various van der Waals heterostructure NDR devices including this work (red and black markers). This plot highlights the temperature sensitivity and performance trends across different 2D material systems. }
    \label{fig:fig5}
\end{figure*}

We now discuss the performance limits of our TFET devices by examining the temperature dependence of the observed NDR in our MoTe$_2$/SnS$_2$ heterojunctions and how it impacts the PVCR values. Figure \ref{fig:fig5}(a) shows the temperature dependence for both the peak (J$_\mathrm{Peak}$) and dip (J$_\mathrm{Dip}$) current densities for Device D2. Correspondingly, the variation of J$_\mathrm{Peak}$ and J$_\mathrm{Dip}$ leads to a certain temperature dependence in PVCR, shown in Fig. \ref{fig:fig5}(b) which is compared to other contemporary van der Waals NDR devices. The maximum experimentally observed PVCR of $\approx$ 4 at T = 150 K in our MoTe$_2$/SnS$_2$ heterostructure is already comparable to other reported van der Waals systems. However, these high values do not persist at either low or high temperatures.

A possible scenario for the low-temperature regime (T $<$ 150 K) is that the reduced phonon population limits phonon-assisted tunneling arising from the momentum mismatch between the valence band maximum of MoTe$_2$ and the conduction band minimum of SnS$_2$ \cite{sriv2018low,fraser2020selective}. The resulting reduction in inelastic tunneling events can suppress carrier transport across the junction. Recent studies on layered TaIrTe$_4$ \cite{du2025evidence} have demonstrated substantial electron-phonon coupling, highlighting the relevance of electron-phonon interactions to the temperature-dependent electronic properties of vdW materials. At lower temperatures, reduced thermal broadening is also expected to decrease the valley current, which can enhance the PVCR. However, the simultaneous reduction in phonon-assisted tunneling and effective DOS overlap may suppress the overall tunneling current, contributing to the observed non-monotonic temperature dependence.

While reduced phonon-assisted processes may limit NDR at lower temperatures, different transport mechanisms become increasingly important as the temperature is raised. In the higher temperature limit (T $>$ 200K), we observe that both J$_\mathrm{Peak}$ and J$_\mathrm{Dip}$ increase; however, the PVCR reduces due to larger increase in the J$_\mathrm{Dip}$ values. This behavior suggests the emergence of additional leakage mechanisms such as thermionic emission or trap-assisted tunneling becoming more prominent at higher temperatures and conceal the NDR signature by increasing the valley current. Moreover, effects like thermal broadening of energy distribution also play significant role in reducing the effective band-to-band-tunneling, thereby further degrading the PVCR. It is in the intermediate temperature regime, T $\sim$ 150 K, the thermally activated carrier injection is optimized while leakage through thermionic and trap-assisted channels remains suppressed, giving rise to a maximum PVCR \cite{huo2024negative,kinoshita2024negative}. 

Compared to other vdW heterostructures, BP-based and graphene-based systems exhibit the highest PVCR values. In BP-based devices, this is largely due to suitability of band alignment engineering, strong anisotropic carrier transport, and high-quality interfaces achieved in homojunctions that minimize leakage and non-resonant tunneling \cite{kim2020thickness,shim2016phosphorene,xiong2020transverse,srivastava2021resonant,yan2015esaki,seo2022van}. In contrast, graphene-based resonant tunneling diodes show high PVCR due to momentum-conserving tunneling across atomically flat barriers like hBN, enabling sharp energy filtering and resonant transmission \cite{britnell2013resonant,fallahazad2015gate,zhang2023toward,liang2025computational,yan2023mote2}. Following such examples, the PVCR in MoTe$_2$/SnS$_2$ heterojunctions could be further enhanced by optimizing band overlap and minimizing parasitic leakage. This could be achieved by for instance, inserting a thin dielectric spacer (e.g., hBN) between semiconducting layers \cite{shim2016phosphorene,kim2024room,xiao2024multifunctional,kinoshita2024negative} to reduced interface disorder and suppress defect-assisted tunneling, which together sharpen the resonance condition and reduce valley current.

\section*{Conclusion}

In summary, we demonstrated a multifunctional dual-gated vdW heterojunction by combining a few layers of MoTe$_2$ and SnS$_2$. The junction exhibits strong gate tunability, transitioning from forward to reverse rectifying behavior at opposite backgate voltages, with a rectification ratio exceeding two orders of magnitude. At intermediate gating conditions, the device exhibits Esaki diode behavior, showing robust negative differential resistance up to temperatures of 200 K. The NDR features are highly tunable via the topgate voltage, with the peak voltage exhibiting a linear dependence on topgate bias (topgate-coupling efficiency $\sim$ 0.5), and a PVCR approaching $\sim$ 4 at 150 K. Using DFT calculations, we find that transport is dominated by valence band-to-valence band tunneling, with an additional valence band-to-conduction band tunneling contribution at higher bias, while gate tunability originates from changes in band overlap and Fermi-level alignment. The gate-tunable features of the NDR curves are attributed to a combined effect of band overlap modulation and Fermi level tuning induced by the topgate. The strong gate coupling enables effective manipulation of broken-gap alignment in the heterostructure. These results establish MoTe$_2$/SnS$_2$ as a versatile platform for exploring tunable tunneling transport and demonstrate how NDR can serve as a sensitive probe of band alignment in 2D heterostructures. Our findings underscore the importance of material choice, band structure engineering, and dual-gating schemes for advancing energy-efficient logic and multifunctional nanoelectronic devices. 

\section*{Acknowledgement}

The authors acknowledge financial support from the 2D TECH VINNOVA Competence Center (No. 2019-00068), the FlagEra projects H2O and MagicTune (funded by VR), the European Union Graphene Flagship project 2D Materials of Future 2DSPIN-TECH (No. 101135853), and the Graphene Center, AoA Nano, AoA Materials, and AoA Energy programs at Chalmers University of Technology. J.W. acknowledges funding from the Swedish Strategic Research Foundation through a Future Research Leader program (FFL21-0129). This work was performed in part at Myfab Chalmers and the Chalmers Materials Analysis Laboratory (CMAL).

\section*{Methods}

We fabricate the heterojunctions using a standard exfoliation and 2D material stacking process. MoTe$_2$, SnS$_2$, and hBN crystals are exfoliated using blue Nitto tape or Scotch tape over small pieces of PDMS sheets. Suitable flakes of the materials are identified and selected under an optical microscope. Using a micro-manipulator equipped with x, y, and z precision stages, as well as rotation and heater configurations, we transfer hBN, MoTe$_2$, and SnS$_2$ onto Si/SiO$_2$ substrates in that order. After the stacking process, the devices are vacuum annealed at 120 $^{\circ}$C for 2 hours at a pressure of approximately 2 $\times$ 10$^{-6}$ mbar. This step helps achieve a residue-free interface.
For contact fabrication, the samples are spin-coated with a bilayer resist followed by e-beam lithography. The source and drain contacts are fabricated in a two-step process, where we pattern the contacts on MoTe$_2$ and SnS$_2$ flakes separately using e-beam lithography, followed by e-gun evaporation of metal and liftoff in hot acetone. We use Pd/Au (15 nm/60 nm) and Ti/Au (5 nm/70 nm) as the metal combinations for MoTe$_2$ and SnS$_2$, respectively. For dual-gated devices, we transfer another hBN layer on top of the junction and define the topgate using a similar patterning process.
All samples are wire-bonded and loaded into a 4K wet cryostat for characterization and measurement. We use a variable gain current amplifier, FEMTO DHPCA-100, for applying biasing voltage and current measurement, and a Keithley 2400 for backgating and topgating purposes.

\section*{Publication note}

This author-prepared manuscript corresponds to the article published in
\textit{Scientific Reports}. It is not the publisher-formatted Version of
Record. The Version of Record is available online at
\href{https://www.nature.com/articles/s41598-026-68365-1}
{https://doi.org/10.1038/s41598-026-68365-1}.

\bibliography{references_sci}

\clearpage
\includepdf[pages=-,
    pagecommand={},
  fitpaper=true
]{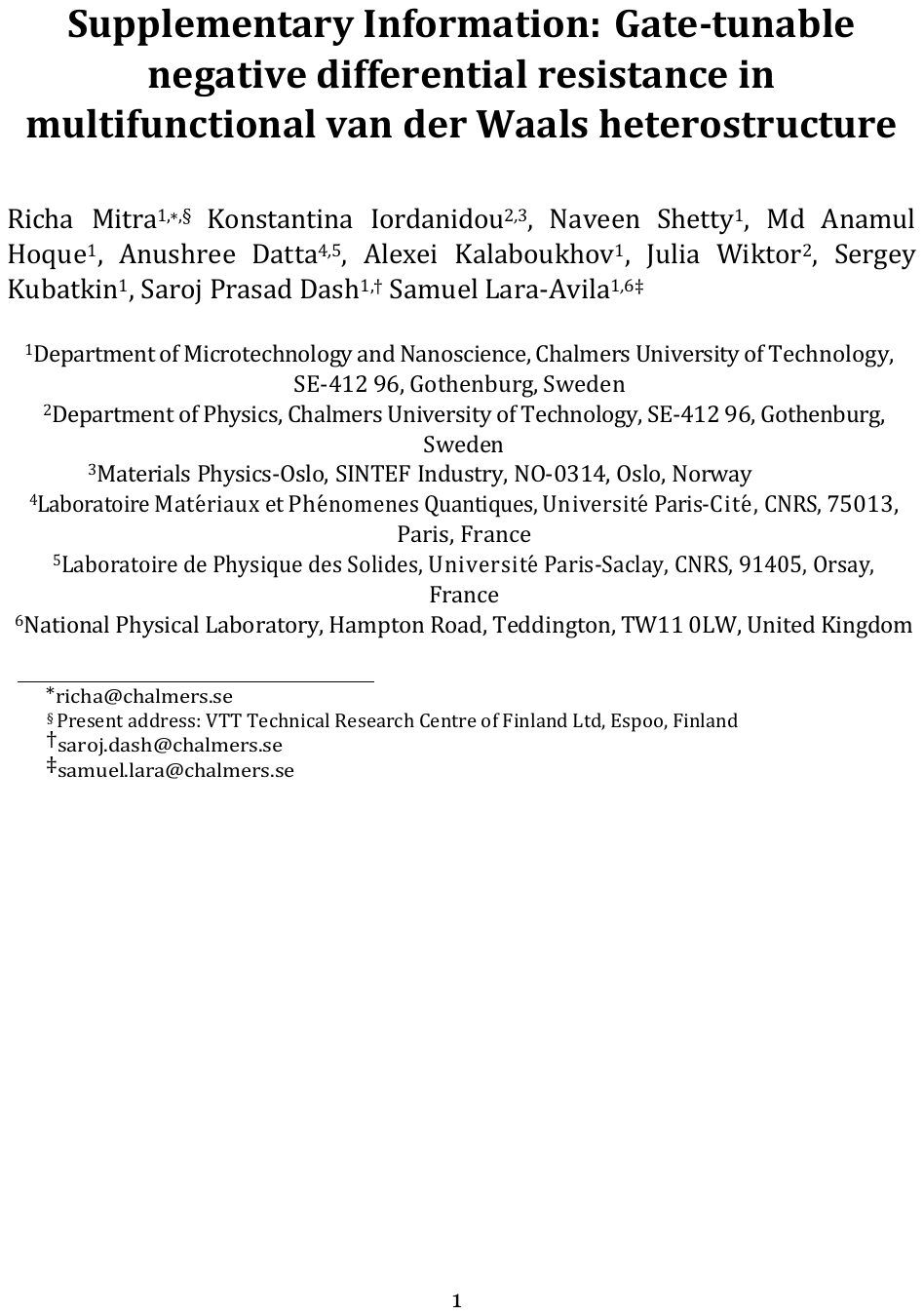}

\end{document}